\documentclass[conference]{IEEEtran}
\IEEEoverridecommandlockouts
\usepackage{cite}
\usepackage{amsmath,amssymb,amsfonts}
\usepackage{algorithmic}
\usepackage{graphicx}
\usepackage{textcomp}
\usepackage{xcolor}
\usepackage{float}

\usepackage{subfigure}

\newtheorem{lemma}{Lemma}

\newtheorem{proposition}[lemma]{Proposition}

\newcommand{\bs}{{\mathbf{s}}}

\newcommand{\bA}{{\textbf{A}}}
\newcommand{\by}{\mathbf{y}}
\newcommand{\bx}{\mathbf{x}}
\newcommand{\bn}{\mathbf{n}}
\newcommand{\bH}{\mathbf{H}}

\newcommand{\bE}{\mathbf{E}}
\newcommand{\bW}{\mathbf{W}}
\newcommand{\bU}{\mathbf{U}}
\newcommand{\bV}{\mathbf{V}}
\newcommand{\bI}{\mathbf{I}}

\newcommand{\bX}{\mathbf{X}}

\newcommand{\tH}{\tilde{\bH}}
\newcommand{\barH}{\bar{\bH}}
\newcommand{\bM}{\mathbf{M}}

\newcommand{\tr}{\textrm{Tr}\,}

\def\BibTeX{{\rm B\kern-.05em{\sc i\kern-.025em b}\kern-.08em
    T\kern-.1667em\lower.7ex\hbox{E}\kern-.125emX}}

\begin{document}

% \title{Conference Paper Title*\\
\title{Learning-Based Massive Beamforming
% \thanks{This work was supported in part by the National Natural
% Science Foundation of China under Grants 61936014 and 61671411, in part by the Fundamental Research Funds for the Central Universities, in part by the National Key Research and Development Project under Grant 2019YFB2102300, 2019YFB2102301 and 2017YFE0119300.}
}

	\author{
	\IEEEauthorblockN{Siyuan Lu, Shengjie Zhao and Qingjiang Shi}
	\IEEEauthorblockA{\textit{School of Software Engineering}\\
	\textit{Tongji University}\\
	Shanghai, China\\
	% \IEEEauthorblockA{\IEEEauthorrefmark{2}School of Automotive Studies\\
	% Tongji University, Shanghai, China}
	% \IEEEauthorblockA{\IEEEauthorrefmark{3}Department of Electronic and Information Engineering\\
	% Tongji University, Shanghai, China}
	Email: \{14\_lusiyuan, shengjiezhao, shiqj\}@tongji.edu.cn
	}}

% \author{\IEEEauthorblockN{1\textsuperscript{st} Siyuan Lu and 3\textsuperscript{rd} Qingjiang Shi}
% \IEEEauthorblockA{\textit{School of Software Engineering} \\
% \textit{Tongji University}\\
% Shanghai, China \\
% \{14\_lusiyuan, shiqj\}@tongji.edu.cn}
% \and
% \IEEEauthorblockN{2\textsuperscript{nd} Shengjie Zhao}
% \IEEEauthorblockA{\textit{School of Software Engineering} \\
% \textit{Tongji University}\\
% \textit{Key Laboratory of Embedded System and Service Computing}\\
% \textit{Ministry of Education}\\
% Shanghai, China \\
% shengjiezhao@tongji.edu.cn}
% \and
% \IEEEauthorblockN{3\textsuperscript{rd} Qingjiang Shi}
% \IEEEauthorblockA{\textit{School of Software Engineering} \\
% \textit{Tongji University}\\
% Shanghai, China \\
% shiqj@tongji.edu.cn}
% \and
% \and
% \and

\maketitle

\begin{abstract}
Developing resource allocation algorithms with strong real-time and high efficiency has been an imperative topic in wireless networks. Conventional optimization-based iterative resource allocation algorithms often suffer from slow convergence, especially for massive multiple-input-multiple-output (MIMO) beamforming problems. This paper studies learning-based efficient massive beamforming methods for multi-user MIMO networks. The considered massive beamforming problem is challenging in two aspects. First, the beamforming matrix to be learned is quite high-dimensional in case with a massive number of antennas. Second, the objective is often time-varying and the solution space is not fixed due to some communication requirements. All these challenges make learning representation for massive beamforming an extremely difficult task. In this paper, by exploiting the structure of the most popular WMMSE beamforming solution, we propose convolutional massive beamforming neural networks (CMBNN) using both supervised and unsupervised learning schemes with particular design of network structure and input/output. Numerical results demonstrate the efficacy of the proposed CMBNN in terms of running time and system throughput.  
\end{abstract}

\begin{IEEEkeywords}
Beamforming, WMMSE, convolutional neural network, massive MIMO
\end{IEEEkeywords}

\section{Introduction}
The rapid development of deep learning in various applications has greatly changed many aspects of our life \cite{simeone2018very}. Besides the changes in human life, many research fields are also revolutionized by deep learning, such as computer vision and natural language processing.
In the research of wireless network communication, deep learning (and machine learning) based methods are gaining more and more attention due to their efficacy. In response to this, embedding deep learning into the 5th generation of mobile systems (5G) and wireless networks is becoming an increasingly hot topic in recent years \cite{sun2018learning,zhang2019deep}.

At the same time, the advantages of massive MIMO in energy efficiency, spectral efficiency, robustness and reliability proved massive MIMO to be indispensable in the 5G era \cite{lu2014overview,larsson2013massive}. To improve the quality of communication in massive MIMO systems, downlink beamforming or precoding is one of the most important transmission technologies. 
% Figure \ref{fig:massive-mimo} shows a typical massive multi-user MIMO beamforming scenario.
For beamforming design, the system throughput (weighted sum-rate) maximization under a total power constraint is an important metric of communication quality, which is the focus of our paper.

Many algorithms developed for beamforming are based on optimization theory like weighted minimum mean square error (WMMSE) \cite{shi2011iteratively}, which can find locally optimal solutions of a formulated optimization problem through iterations.
However, such optimization based algorithms often suffer from high computational costs (e.g., WMMSE involves complex matrix inversion operations).
When large-scale antenna arrays are deployed on transmitter \cite{vook2014mimo}, the computational cost of these algorithms can be prohibitive. Meanwhile, algorithms with low complexity like zero-forcing method \cite{yoo2006optimality} cannot achieve good performance when the number of users or antennas becomes large.
% such methods with high complexity can not meet the needs of 5G communication.
% As a result, deep learning-based methods like BNNs \cite{xia2019deep} were proposed to solve such problems recently. And deep learning-based optimization was proven to be greatly useful for improving the efficiency of large scale network optimization \cite{liu2017deep} in network communication. 

As a result, deep learning-based methods were proposed to solve such problems in recent years. Supervised deep neural network (DNN) has been applied to power control, which can achieve similar sum-rate performance as the classical power allocation algorithm WMMSE \cite{sun2018learning}. In contrast, unsupervised learning can reach (even better) the performance of the WMMSE algorithm \cite{huang2018unsupervised,liang2018towards}. Meanwhile, a hybrid precoding scheme with DNN-based autoencoder \cite{huang2019deep} was proposed for Millimeter wave (mmWave) MIMO systems. Apart from the aforementioned DNN models, a distributed convolutional neural network (CNN)-based deep power control network was introduced \cite{lee2018deep} to maximize the system spectral efficiency or energy efficiency with local CSI. Furthermore, CNN-based beamforming neural networks (BNNs) were proposed \cite{xia2019deep} for three typical beamforming optimization problems in multi-user multiple-input-single-output (MISO) networks. For the sum-rate maximization problem, BNNs were trained using both supervised learning and unsupervised learning. 

Although these deep learning-based methods have been proposed for multi-user MIMO downlink beamforming, current methods mainly focus on the basic case of the sum-rate maximization problem without taking more complicated situations like user priority or varying number of stream per user into consideration. Besides, the neural network design does not utilize the structure of the closed-form update in the iterative algorithm. 

In summary, two main challenges remain unaddressed in learning-based massive MIMO beamforming. First, as the number of antennas becomes large in massive MIMO system, both the input and output of the neural network (NN)-based methods would be of high dimension, which makes the neural network more complex and harder to train.
Second, in real-world systems, the number of user streams and user priority are both changeable over time which means the solution space is not fixed. Thus, it will be quite challenging to take such two cases into consideration without increasing the neural network complexity significantly.

In this paper, to tackle the above challenges, we propose a new deep learning framework called convolutional massive beamforing neural networks (CMBNN). The main contributions of this paper are summarized as follows: 

1) We utilize the structure of the closed-form solution of WMMSE algorithm in the design of the NN structure. In addition, we design a novel NN structure to cope with varying number of user streams. By doing so, for the first time, we are able to handle beamforming with time-varying user priority and varying number of user streams without significantly increasing the NN complexity or sacrificing model performance. 

2) Due to the use of problem structure in the design of our networks, all NN structures proposed in our paper are much simpler than existing approaches. The low complexity of NN structures makes our method more appealing under the real-time requirements in 5G wireless communication systems.

\section{System Model and Problem Formulation}
\label{sec:formulation}
\subsection{System Model}
Consider a single cell multi-user massive MIMO system where the BS is equipped with $N_T$ transmit antennas and serves $K$ users each equipped with $N_R$ antennas \cite{caire2003achievable}. Let $\bV_{k} \in \mathbb{C}^{N_T \times d_{k}}$ denotes the transmit beamforming that the BS employs to send the signal $\bs_{k} \in \mathbb{C}^{d_{k} \times 1}$ to user $k$. The BS signal is given by,
\begin{align}
\bx = \sum_{k=1}^{K} \bV_{k} \bs_{k}, \nonumber
\end{align}
 % $\bx = \sum_{k=1}^{K} \bV_{k} \bs_{k}$ 
where it is assumed $\mathbb{E}\left[\bs_{k}\bs_{k}^H\right]= \bI$.

Assuming a flat-fading channel model, the received signal $\by_{k}\in \mathbb{C}^{N_{R}\times1}$ at user $k$ can be written as
\begin{align}
\by_{k} &= \bH_k\bx+\bn_k\\
&=\underbrace{\bH_{k}\bV_{k}\bs_{k}}_{\textrm{desired signal of user $k$}}+\underbrace{\sum_{j=1,j \neq k}^K\bH_{k}\bV_{j}\bs_{j}}_{\textrm{multi-user interference}}+\bn_{k}, \forall k\nonumber
\end{align}
where matrix $\bH_{k}\in\mathbb{C}^{N_{R} \times N_T}$ represents the channel matrix from the BS to user $k$, while $\bn_{k} \in \mathbb{C}^{N_R\times 1}$ denotes the additive white Gaussian noise with distribution $\mathcal{CN}(0,\sigma_{k}^2 \bI)$. We assume that the signals for different users are independent from each other and from receiver noises. In this paper, we treat the multi-user interference as noise and employ linear receive beamforming strategy, i.e., $\bU_{k} \in \mathbb{C}^{d_k \times N_{R}}, \forall k$, so that the estimated signal $\hat{\bs}_{k}\in  \mathbb{C}^{d_k \times 1} $ is given by $\hat{\bs}_{k} = \bU_{k}^H \by_{k},  \forall k.$

\subsection{Problem Formulation}
A basic problem of interest is to find the transmit beamformers $\{\bV_{k}\}$ such that the system weighted sum-rate is maximized subject to a total power constraint due to the BS power budget. Mathematically, it can be written as follows
\begin{equation}\label{eq:sum_rate_MIMO}
\begin{split}
\max_{\{\bV_k\}} \quad &\sum_{k=1}^K \alpha_{k}R_{k}\\
\textrm{s.t.}\quad &\sum_{k=1}^{K}\tr(\bV_{k}\bV_{k}^H)\leq P_{max}
\end{split}
\end{equation}
where $P_{max}$ denotes the BS power budget,  the weight $\alpha_{k}$ represents the priority of user~$k$ in the system, and $R_{k}$ is the rate of user $k$ given by
\begin{equation}\label{eq:R}
R_{k} \triangleq \log\det\left(\bI{+}\bH_{k}\bV_{k}\bV_{k}^H\bH_{k}^H \left(\bA_k{+}\sigma_{k}^2\bI\right)^{-1}\right).\nonumber
\end{equation}
where $\bA_k \triangleq \sum_{j\neq k}\bH_{k}\bV_{j}\bV_{j}^H\bH_{k}^H$.

Under the independence assumption of $\bs_{k}$'s and $\bn_{k}$'s, the MSE matrix $\bE_{k}$ can be written as,
\begin{equation}\label{eq:MSE}
\begin{split}
\bE_{k} & \triangleq (\bI-\bU_{k}^H \bH_{k} \bV_{k})(\bI- \bU_{k}^H \bH_{k}\bV_{k})^H \\
	& + \sum_{m\neq k} \bU_{k} \bH_{k} \bV_{m}\bV_{m}^H \bH_{k}^H \bU_{k}^H \\
	& + \sum_{i=1}^{K}\frac{\sigma_k^2}{P_{max}}\tr(\bV_{i}\bV_{i}^H) \bU_{k}^H \bU_{k}
\end{split}
\end{equation}

Followed by \cite{shi2011iteratively}, we can obtain the equivalent WMMSE form as 
\begin{equation*}
	\min_{\{\bW_k, \bU_k, \bV_k\}}\quad \sum_{k=1}^K\left(\log\det(\bW_{k})-{\rm \tr}(\bW_{k} \bE_{k})\right)
\end{equation*}

Furthermore, inspired by the structure of ZF beamforming \cite{yoo2006optimality}, to reduce the complexity in the massive antenna scenario, we also restrict $\bV_k$'s  to the range space of $\bH^H$, i.e., let it satisfy $\bV_k=\bH^H\bX_k$ with some $\bX_k\in\mathbb{C}^{KN_{R} \times d_{k}}$, where $\bH\triangleq\left[\bH_1^H ~~\bH_2^H ~~\ldots \bH_K^H\right]^H\in\mathbb{C}^{KN_{R} \times N_{T}}$.  As a result, by defining $\bM_k\triangleq\bU_k\bW_k\bU_k^H$ and $\barH\triangleq \bH\bH^H\in\mathbb{C}^{KN_R\times KN_R}$, we can derive the three main steps of the corresponding WMMSE algorithm as follows
\begin{align}\label{eq:X}
\bX_k &= \left(\sum_{j=1}^K \frac{\sigma_k^2}{P_{max}}\alpha_j\tr(\bM_j)\barH+\sum_{i=1}^K \alpha_i\barH_i^H\bM_i\barH_i \right)^{-1}\nonumber\\
	&~~~~ \times \alpha_k\barH_k^H\bU_k\bW_k\\
\bU_k &= \left(\sum_{j=1}^K \frac{\sigma_k^2}{P_{max}}\tr(\barH\bX_j\bX_j^H) \bI+\sum_{i=1}^K \barH_i\bX_i\bX_i^H\barH_i^H \right)^{-1}\nonumber\\
&~~~~\times \barH_k\bX_k\\
\bW_k &= \left(\bE_k \right)^{-1} = \left(\bI{-}\bU_k^H\barH_k\bX_k \right)^{-1}
\end{align}
The algorithm repeats the above three steps until convergence. For ease of exposition, it is termed as reduced-WMMSE (R-WMMSE).

\section{Proposed Method}
\label{sec:method}
Our key idea is to learn the R-WMMSE algorithm above using deep learning, so that the complexity can be further reduced by choosing appropriate neural network structure and input/output.

\subsection{Reformulation}
% Basically, the NN input could include the channel matrices and noise power. 
In previous work like \cite{xia2019deep}, the noise power $\sigma_k^2$ is often fixed for all scenarios, resulting in the trained network only adapting to this noise level. Here we remove the effect of noise by reformulating the problem. Let us define $\tH_k=\sqrt{\frac{P_{max}}{\sigma_k^2}}\bH_k$ and 
\begin{align}
\tilde{R}_k\!\triangleq\! \log\det\!\Bigg(\!\bI{+}\tH_{k}\bV_{k}\bV_{k}^H\tH_{k}^H\nonumber \!\left.\left(\sum_{j\neq k}\tH_{k}\bV_{j}\bV_{j}^H\tH_{k}^H\!{+}\bI\!\right)^{-1}\!\right)
\end{align}
Then we have the following proposition.
\begin{proposition}
	Problem \eqref{eq:sum_rate_MIMO} is equivalent to 
	\begin{equation}\label{eq:sum_rate_MIMO2}
	\begin{split}
	\max_{\{\bV_k\}} \quad &\sum_{k=1}^K \alpha_{k}\tilde{R}_k\\
	\textrm{s.t.}\quad &\sum_{k=1}^{K}\tr(\bV_{k}\bV_{k}^H)\leq 1,
	\end{split}
	\end{equation}
in the sense that the optimal solution to problem \eqref{eq:sum_rate_MIMO2} multiplied by $\sqrt{P_{max}}$ is also optimum to  problem \eqref{eq:sum_rate_MIMO}. 
\end{proposition}
% According to the result of Proposition 1, we  perform data preprocessing of the channels, i.e., let $\tH_k=\sqrt{\frac{P_{max}}{\sigma_k^2}}\bH_k$, which includes not only the effect of channels but also the effect of the noise power and the BS power budget. 
Because of the above equivalence,  we consider problem \eqref{eq:sum_rate_MIMO2} throughout the rest of this paper. Moreover, for notational simplicity, we drop `~$\tilde{}$~' in all notations in \eqref{eq:sum_rate_MIMO2}.

\subsection{Neural Network Architecture}
\begin{figure}[b]
	\centering
	\includegraphics[width=0.46\textwidth]{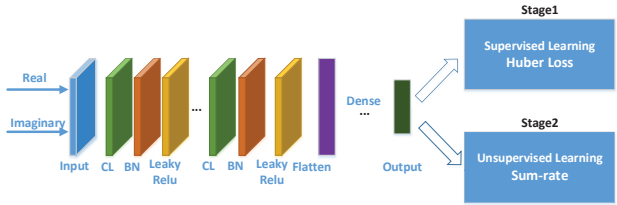}
	\caption{A basic neural network structure for massive beamforming}
	\label{fig:Supervised+Unsupervised}
\end{figure}

Figure \ref{fig:Supervised+Unsupervised} presents the CNN-based network architecture for beamforming design followed by the idea of \cite{xia2019deep}, where CL and BN denote the convolutional layer and batch normalization layer respectively, leaky relu is chosen as the activation function and several dense layers are used after the flatten layer. The network architecture is further detailed as follows.
\subsubsection{Supervised Learning}
Supervised learning is a straightforward way to train a beamforming neural network. All data samples can be generated through running the R-WMMSE algorithm. For the CNN model, as the input $\bH$ or $\bH\bH^{H}$ are all complex matrices, we would like to reshape the complex matrix to a tensor like an image but with only two channels, one represents the real part while the other represents the imaginary part. However, different from the traditional image processing with convolutional and pooling layers, we would not use pooling layer because it may cause information loss which would influence the learning result. Adam and huber loss are selected as the optimizer and loss function respectively. 

\subsubsection{Unsupervised Learning}
Even if the huber loss of supervised learning becomes small, the weighted sum-rate result is not necessarily large enough. The intuitive reason is that the supervised learning does not aim directly to maximize the weighted sum-rate and its performance is largely limited by the training samples. On the other hand, we have a direct objective, i.e., weighted sum-rate maximization. Hence, we could use the negative weighted sum-rate as an alternative training loss which could improve the sum-rate directly.
	\begin{equation}
		L(\theta; h) \triangleq -\sum_{k=1}^{K} \alpha_{k} R_k(h, o)
	\end{equation}

\subsubsection{Supervised + Unsupervised Learning}
As the loss function of unsupervised learning is complicated involving many complex matrix operations, both the loss calculation and the corresponding gradient computation would be more time-consuming than the computation of traditional loss (e.g., MSE). Considering the trade-off between convergence speed and accuracy, we choose to combine both supervised learning and unsupervised learning to train the beamforming neural network. % solution to the weighted sum-rate maximization problem. 
Specifically, supervised learning is used for pre-training and unsupervised learning is for further refinement. In practice, only one or two epochs for unsupervised learning is enough.

\subsection{Design of Input and Output}
In massive MIMO system, the number of transmit antennas $N_T$ could be very large. Hence, if we still directly take $\bH$ and $\bV_{k}$ (or $\bX_k$) as input and output as in \cite{xia2019deep,huang2018unsupervised}, the input/output of the neural network (NN) would be both high dimensional matrices, making it not easy to train an NN. As a consequence, the NN input and output should be redesigned to reduce the NN input/output size (and thus the training complexity and difficulty).
In terms of the R-WMMSE algorithm, we find that beamformer $\bV_{k}$ is uniquely determined by $\bX_{k}$ while $\bX_{k}$ depends on $\bH\bH^H$. Hence, $\bH\bH^H$ can be regarded as the NN input, which has reduced size as compared to $\bH$ when $N_T>>N_R$. Moreover, since $\bX_{k}$ can be determined by $\bH\bH^H$ and $\{\bU_k, \bW_k\}$, we can take $\{\bU_k, \bW_k\}$ as the NN output in order to reduce the size of NN output. Tables \ref{table1} and \ref{table2} list the dimension of various input/output schemes. It can be seen that different choice of input/output leads to different size of input/output. Note that we generally have $N_T>>N_R$, $N_T\geq KN_R$, $K\geq N_R, d_k$ in the massive MIMO case.
\begin{table}[!htbp]
		\centering
		\caption{Dimension of different inputs}
		\begin{tabular}{c c}
			\hline
			Input & Dimension\\
			\hline
			$\bH_k$ & $2 \times (KN_{R} \times N_{T})$\\
			$\bH\bH^H$ & $2 \times (KN_{R} \times KN_{R})$\\
			% $\bH\bH^H$(Conjugate Symmetry) & $KN_{R} \times KN_{R}$\\
			\hline
		\end{tabular}
		\label{table1}
\end{table}

\begin{table}[!htbp]
		\centering
		\caption{Dimension of different outputs}
		\begin{tabular}{c c}
			\hline
			Output & Dimension\\
			\hline
			$\bV_k$ & $2 \times (N_T \times d_{k})$\\
			$\bX_k$ & $2 \times (KN_{R} \times d_{k})$\\
			$\bU_k$ and $\bW_k$ & $2 \times (N_{R} \times d_{k} + d_{k}\times d_{k})$\\
			% $\bU_k$ and $\bW_k$ (Conjugate Symmetry) & $2 \times (N_{R} \times d_{k}) + d_{k}\times d_{k}$\\
			\hline
		\end{tabular}
		\label{table2}
\end{table}

\begin{figure}[!htbp]
	\centering
	\includegraphics[width=0.32\textwidth]{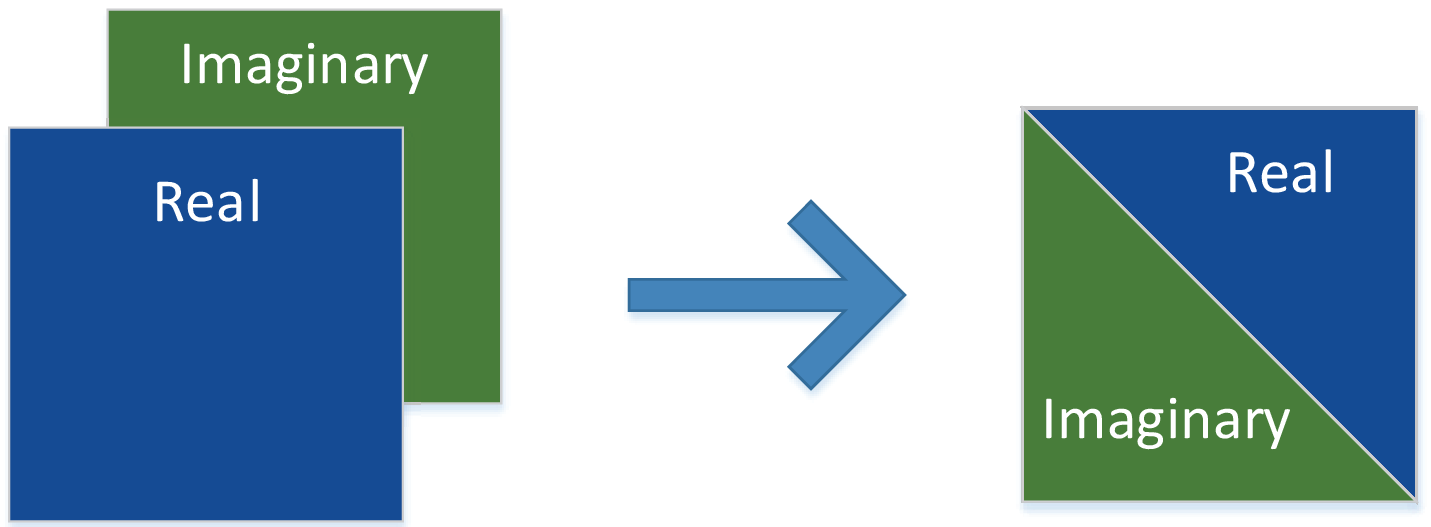}
	\caption{Reconstruction of input $\bH\bH^H$}
	\label{fig:new_HH}
\end{figure}

Furthermore, due to the conjugate symmetry of $\bH\bH^H$, 
both the real part and imaginary part of $\bH\bH^H$ depend uniquely on their upper or lower triangular parts. Hence, we can further reduce the input size by combining the real part and the imaginary part in a way as shown in Fig. \ref{fig:new_HH}.  As a result, the dimension of NN input is finally reduced to $KN_{R} \times KN_{R}$. Similar operation can be done for $\bW_k$, leading to a further reduced size of NN output.

\subsection{Architecture Design for User Priority}
In practice, each user $k$ in the system may have a different priority with weight $\alpha_k$ that often changes with time. While most current  methods do not take this into consideration, the NN input or structure should be carefully redesigned when the weights are considered.
% Considering the changeable weight $\alpha_k$, the network input or the network structure should be carefully redesigned. 
According to the R-WMMSE algorithm mentioned before, both $\bX_k$ and $\{\bU_k,\bW_k\}$ depend on $\alpha_k\bH\bH^H$. 

\begin{figure}[t]
	\centering
	\includegraphics[width=0.2\textwidth]{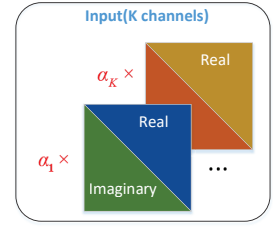}
	\caption{Merge weight into input as K channels}
	\label{fig:Weight_Channels}
\end{figure}

\begin{figure}[t]
	\centering
	\includegraphics[width=0.36\textwidth]{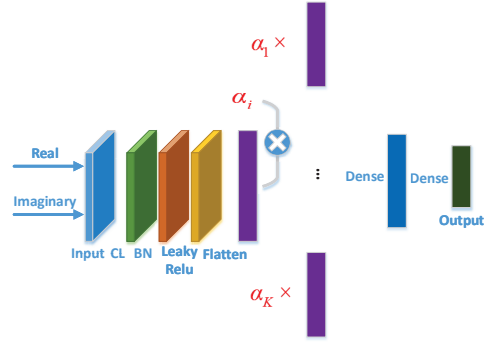}
	\caption{Concatenate weight after conv.}
	\label{fig:Weight_Merge}
\end{figure}

Two very intuitive ways to merge the weights into the NN are depicted in Figure \ref{fig:Weight_Channels} and Figure \ref{fig:Weight_Merge}. One is to merge weights into input as K channels (see Figure \ref{fig:Weight_Channels}), and the other is to concatenate weight after convolution and flatten of the input (see  Figure \ref{fig:Weight_Merge}). Our simulation results show that these two methods can achieve reasonably good performance.% as that in the simple case without considering the user priority.

However, these two methods will bring higher computational complexity to the original network which can lead to extra time and cost. Surprisingly, inspired by the update rule of $\bX_k$ in \eqref{eq:X}, we find that, just by taking $\tH\tH^H$ as input, where $\tH_{k} = \sqrt\alpha_k\bH_k$ and $\tH\triangleq\left[\tH_1^H ~~\tH_2^H ~~\ldots \tH_K^H\right]^H$,
we can reach the same performance as the previous two intuitive methods without any need for increasing network complexity.
% Simulation result shows that it can achieve similar performance like such two methods and is much more simple.

\subsection{Architecture Design for varying number of user streams}
In practical systems, sometimes only one stream is transmitted for some user during communication. This raises a new challenge that the number of streams $d_k \ (d_k \leq N_R)$ can vary but the dimension of the network output needs to be fixed. Table \ref{table3} shows the number of valid output elements when $d_k$ is different. Thus, to ensure that the network output have fixed dimension, certain positions should be set to zero when there exists a single stream transmission.
 % which means some parts of the output $\bU_k$ and $\bW_k$ should be zeroed. 
\begin{table}[!htbp]
		\centering
		\caption{Dimension of different $d_k$}
		\begin{tabular}{c c c c}
			\hline
			$d_k$ & $\bU_k$ & $\bW_k$ & Num of valid elements\\
			\hline
			2 & $2\times 2$ & $2\times 2$ & 12\\
			% \hline
			1 & $2\times 1$ & $1\times 1$ & 5\\
			\hline	
		\end{tabular}
		\label{table3}
\end{table}

\begin{figure}[!htbp]
	\centering
	\includegraphics[width=0.42\textwidth]{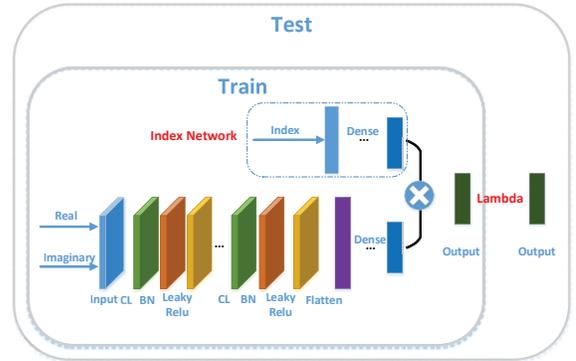}
	\caption{Network structure for varying number of user streams}
	\label{fig:Index_Merge}
\end{figure}

There are a few simple and intuitive solutions to this problem. 
The simplest solution is to directly merge the information of the number of user streams into the original input $\bH\bH^H$.
% The simplest solution is to directly add the number of streams as extra input in parallel with $\bH\bH^H$ to the network.
Another solution is to ignore the number of streams used and manually set to zero the positions corresponding to empty streams of the output,  which may result in discontinuous loss function. These two methods both result in unsatisfactory performance in our experiments.

To achieve better performance than the solutions mentioned above, we introduce an auxiliary network called \emph{Index Network} whose function is to softly zero out the corresponding positions given the number of streams used.
Specifically, as shown in Figure \ref{fig:Index_Merge}, the \textit{Index Network} is a network separated from the main network, it takes the number of user streams as input and outputs a soft mask having the same dimension of the output of the main network. The output of the main network is multiplied by the mask element-wisely to produce the final output.  We find that this method can effectively stabilize the training and improve the performance. 

During the testing stage, to ensure that the network outputs a beamforming with correct number of streams, the output elements at certain positions will be set to zero manually at the last layer (Lambda Layer).

During the unsupervised learning phase, variables of the \emph{Index Network} should be fixed and specific positions should also be assigned with zero before calculating the unsupervised loss.

\section{Experiments}
\subsection{Neural Network configuration}
The main network consists of one convolutional layer with 4 kernels of size $3\times3$, followed by batch normalization (BN) layer and activation function layer (leaky relu), then only one dense layer with 32 hidden units. The \textit{Index Network} is of similar scale as the dense layer before. Our network is much simpler than the previous work \cite{xia2019deep,huang2018unsupervised} with much more layers and hidden units (mostly having more than two convolutional and dense layers).

\subsection{Data generation}
For weighted sum-rate maximization, the channel matrix $\bH$ is generated from the complex Gaussian distribution with pathloss between the users and the BS. The pathloss is set to $128.1+37.6\log_{10}(\omega)$[dB] \cite{dahrouj2010coordinated} where $\omega$ is the distance between the user and the BS in km ($0.1\sim0.3$). The noise power is set to be the same for all users and can be calculated by $\sigma_{k}^2 = 10^{\frac{1}{K} \sum_{k} \log_{10}\frac{1}{N_R} \sum_{i,j}\bH_{k_{ij}}^2} \times 10^{-\frac{\rm SNR}{10}}$, where signal-to-noise ratio (SNR) is set as $20 {\rm (dB)}$. The priority coefficients $\alpha_{k}$ of the users are generated randomly and $\sum_{k} \alpha_{k} = K$, and $d_{k}$ is also generated randomly for each user $k$ ($d_{k} = 1$ indicates dual stream and $d_{k} = 0$ indicates single stream).
In the simulation, 45000 samples are generated for training and 5000 are for testing.
Table \ref{table4} lists three main test cases in the following experiment with $N_R = 2$. The last test case is of great importance in industry. 
\begin{table}[!htbp]
		\centering
		\caption{Three main beamforming test cases}
		\begin{tabular}{c c c}
			\hline
			Case & $N_T$ & $K$\\
			\hline
			1 & 8 & 2\\
			% \hline
			2 & 8 & 4\\
			3 & 32 & 12\\
			\hline	
		\end{tabular}
		\label{table4}
\end{table}

\subsection{Simulation Result}
To test whether the predicted precoder $\bV$ is good enough to maximize the weighted sum-rate maximization problem, the predicted output should be put back to the objective function and the performance can be defined as follows.
 % which means the sum-rate value is the actual metric.

\begin{equation}
	f(\bH, \bV_k) \triangleq \sum_{k=1}^{K} \alpha_{k}\tilde{R}_k(\bH, \bV_k)
\end{equation}

\begin{equation}
	Performance \triangleq \frac{f(\bH, \bV_{predict})}{f(\bH, \bV_{true})}
\end{equation}

\begin{figure}[t]
	\centering
	\includegraphics[width=0.4\textwidth]{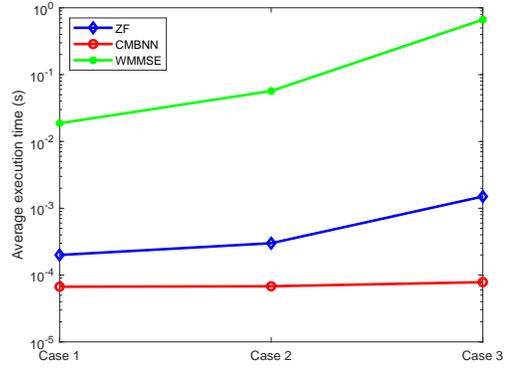}
	\caption{Average execution time (in seconds) among CMBNN, ZF and WMMSE}
	\label{fig:running_time}
\end{figure}

\begin{figure}[t]
	\centering
	\includegraphics[width=0.4\textwidth]{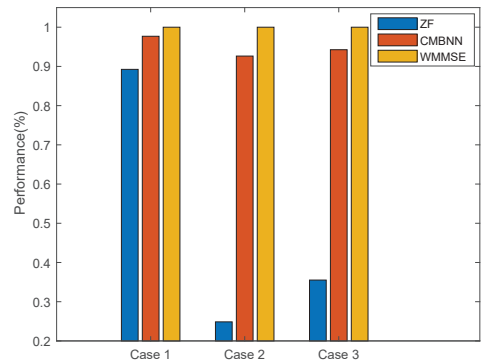}
	\caption{Average performance among CMBNN, ZF and WMMSE}
	\label{fig:Performance}
\end{figure}

Figures \ref{fig:running_time} and \ref{fig:Performance} show the average execution time and average performance compared with the WMMSE algorithm and ZF algorithm respectively. It can be observed that 1) the proposed method can achieve similar performance as the WMMSE algorithm in most cases, while significantly outperforming ZF algorithm (which is unable to handle the user priority); and 2) the proposed method costs less execution time (in testing stage) than both the WMMSE method including many inversion operations and ZF method which needs extra time to decide which beamforming vector should be used for sending single stream.% should be left.

In summary, our proposed CMBNN model is superior from the perspective of both performance and efficiency.

\section{Conclusion}
In this paper, we have proposed a convolutional massive beamforming neural networks (CMBNN) with low complexity. Specifically, we have designed the neural network according to the structure of optimization problem to handle complex situations with changeable user priority and varying number of user streams. Compared with the methods in literature, our proposed framework can achieve better performance and higher efficiency.

\bibliographystyle{IEEEtran}
\bibliography{mybibfile}

\end{document}